\documentclass[prd,superscriptaddress,amsfonts,amssymb,amsmath,showpacs]{revtex4-2}
\usepackage{bm}
\usepackage{amsfonts}
\usepackage{latexsym}
\usepackage[latin1]{inputenc}
\usepackage{graphicx}
\usepackage{amsmath}
\usepackage{palatino}
\usepackage{mathpazo}
\usepackage{textcomp} 
\usepackage{float}
\usepackage{booktabs}
\usepackage{dcolumn}
\usepackage{ragged2e}
\usepackage{hyperref}
\hypersetup{colorlinks,citecolor=green}
\hypersetup{colorlinks=true,linkcolor=blue,filecolor=blue,    urlcolor=magenta}
\usepackage{amsmath}
\usepackage{xcolor}
\usepackage{orcidlink}
\usepackage[caption=false]{subfig}
\usepackage{commath}
\usepackage[autostyle]{csquotes}
\def\jnl@style{\it}
\def\aaref@jnl#1{{\jnl@style#1}}

\def\aaref@jnl#1{{\jnl@style#1}}

\def\aj{\aaref@jnl{AJ}}                   % Astronomical Journal
\def\apj{\aaref@jnl{ApJ}}                 % Astrophysical Journal
\def\apjl{\aaref@jnl{ApJ}}                % Astrophysical Journal, Letters
\def\apjs{\aaref@jnl{ApJS}}               % Astrophysical Journal, Supplement
\def\apss{\aaref@jnl{Ap\&SS}}             % Astrophysics and Space Science
\def\aap{\aaref@jnl{A\&A}}                % Astronomy and Astrophysics
\def\aapr{\aaref@jnl{A\&A~Rev.}}          % Astronomy and Astrophysics Reviews
\def\aaps{\aaref@jnl{A\&AS}}              % Astronomy and Astrophysics, Supplement
\def\mnras{\aaref@jnl{Mon.~Not.~Roy.~Astron.~Soc.}}             % Monthly Notices of the RAS
\def\prd{\aaref@jnl{Phys.~Rev.~D}}        % Physical Review D
\def\plb{\aaref@jnl{Phys.~Lett.~B}}        % Physics Letters B
\def\prc{\aaref@jnl{Phys.~Rev.~C}}  % Physical Review C
\def\prl{\aaref@jnl{Phys.~Rev.~Lett.}}    % Physical Review Letters
\def\qjras{\aaref@jnl{QJRAS}}             % Quarterly Journal of the RAS
\def\skytel{\aaref@jnl{S\&T}}             % Sky and Telescope
\def\ssr{\aaref@jnl{Space~Sci.~Rev.}}     % Space Science Reviews
\def\zap{\aaref@jnl{ZAp}}                 % Zeitschrift fuer Astrophysik
\def\nat{\aaref@jnl{Nature}}              % Nature
\def\aplett{\aaref@jnl{Astrophys.~Lett.}} % Astrophysics Letters
\def\apspr{\aaref@jnl{Astrophys.~Space~Phys.~Res.}} % Astrophysics Space Physics Research
\def\physrep{\aaref@jnl{Phys.~Rep.}}      % Physics Reports
\def\physscr{\aaref@jnl{Phys.~Scr}}       % Physica Scripta
\def\commat{\aaref@jnl{Comm.~Math.~Phys.}}              % Communications in Mathematical Physics
\def\science{\aaref@jnl{Science}}               % Science
\def\cqg{\aaref@jnl{Classical Quant.~Grav.}}            % Classical and Quantum Gravity
\def\jpcs{\aaref@jnl{JPCS}}                                     % Journal of Physics Conference Series
\def\ijmpd{\aaref@jnl{Int.~J.~Mod.~Phys.~D}}                    % International Journal of Modern Physics D
\def\grg{\aaref@jnl{Gen.~Relat.~Gravit.}}               % General Relativity and Gravitation
\def\rpp{\aaref@jnl{Rep.~Prog.~Phys.}}          % Reports on Progress in Physics
\def\npa{\aaref@jnl{Nucl.~Phys.~A}}        % Nuclear Physics A
\def\lrr{\aaref@jnl{Living Rev.~Rel.}}                   % Living reviews in relativity
\def\jcap{\aaref@jnl{J.~Cosmology Astropart.~Phys.}}    % Journal of cosmology and astroparticle physics
\def\rmp{\aaref@jnl{Rev.~Mod.~Phys.}}   %Reviews of modern physics
\def\epjc{\aaref@jnl{Eur.~Phys.~J.~C}}

\allowdisplaybreaks[1]
\begin{document}

\color{black}       %% For one column

\title{Non-Commutative Effects on Wormholes in Rastall-Rainbow Gravity }

\author{Anirudh Pradhan} %\orcidlink{0000-0002-1932-8432}}
\email[]{pradhan.anirudh@gmail.com}
\affiliation{Centre for Cosmology, Astrophysics and Space Science, GLA University, Mathura-281 406, Uttar Pradesh, India}

\author{Safiqul Islam} \email[]{sislam@kfu.edu.sa}
\affiliation{Department of Basic Sciences, Deanship of Preparatory Year, King Faisal University, Hofuf 31982, Al-Hasa, Saudi Arabia}

\author{M. Zeyauddin} %\orcidlink{0000-0001-8382-8994}}
\email[]{uddin\_m@rcjy.edu.sa}
\affiliation{Department of General Studies (Mathematics)
Jubail Industrial College, Jubail 31961, Saudi Arabia} 

\author{Ayan Banerjee} %\orcidlink{0000-0003-3422-8233}} 
\email{ayanbanerjeemath@gmail.com}
\affiliation{Astrophysics and Cosmology Research Unit, School of Mathematics, Statistics and Computer Science, University of KwaZulu--Natal, Private Bag X54001, Durban 4000, South Africa}

%%%%%%%%%%%%%%%%%%%%%%%%%%%%%%%%%%%%%  DATE  %%%%%%%%%%%%%%%%%%%%%%%%%%%%%%%%%%%%

\date{\today}

\begin{abstract}
In this present article, we explore the physical properties and characteristics of static, spherically symmetric
wormholes in the background of Rastall-Rainbow gravity. The Rastall-Rainbow gravity theory has recently been proposed 
as a combination of two theories, namely, the Rastall theory and the Rainbow description. We implemented noncommutativity by adopting two different distributions of energy density (Gaussian and Lorentzian) in the Morris and Thorne metric. We solve the field equations analytically and discuss all the properties of wormholes depending on the two  model parameters. Notably, for specific parameter ranges, one can alleviate the violation of the WEC at the throat and its neighbourhood.

\end{abstract}

\maketitle

\section{Introduction}

Wormholes represent a captivating theoretical concept arising from the general theory of relativity (GR). These hypothetical structures resemble handles and serve as hypothetical bridges linking distant regions within our universe or even connecting different universes. The historical origins of wormhole predictions trace back to the work of Austrian physicist Ludwig Flamm in 1916, shortly after Albert Einstein proposed GR. Nearly two decades later, Albert Einstein and Nathan Rosen \cite{Einstein:1953tkd} introduced the notion of a bridge linking asymptotic regions within a two-sided Schwarzschild black hole, a concept known as the Einstein-Rosen bridge. In 1988,  Morris and Thorne \cite{Morris:1988cz} put forth the idea of a static traversable wormhole, which could connect two asymptotically flat spacetimes. This pioneering work laid the foundation for a traversable wormhole, offering potential for swift interstellar travel by advanced civilizations. Subsequently, Morris, Thorne, and  Yurtsever \cite{Morris1988} explored the possibility of transforming a wormhole into a time machine. Collectively, these seminal contributions have propelled wormholes into a dynamic field of research within the realms of general relativity and alternative theories of gravitation. Interested researchers can delve deeper into Lorentzian wormholes by consulting additional references \cite{Visser:1995,Lobo:2007zb} for comprehensive reviews on the subject.

Many believe that delving into aspects of quantum gravity can be more effectively pursued through mathematical exploration using non-commutative geometry. This belief is founded on the inherent non-commutativity of coordinates, as encapsulated in the commutator, [$x_\mu, x_\nu]=\theta_{\mu \nu}$, where $\theta_{\mu \nu}$ represents an anti-symmetric, real, second-order matrix that plays a pivotal role in defining the fundamental cell discretization of spacetime \cite{Doplicher:1994,Kase:2003,Smailagic:2004,Nicolini:2009}. The authors in \cite{Rahaman:2015} explore the creation of novel wormhole solutions, by introducing an anisotropic real matter source, where the background geometry is influenced by noncommutativity and utilizes conformal Killing vectors to impose constraints on the metric tensor's structure. The unique feature of this noncommutative geometry lies in its ability to replace the point-like gravitational source with a more diffuse energy density distribution, which follows a Gaussian distribution. Non-commutative geometry within the context of f(R) gravity is studied in \cite{Baruah:2023} where the feasibility of constructing wormholes that adhere to various energy conditions, even when employing a source akin to phantom matter, is demonstarted. Through a meticulous examination utilizing well-defined model parameters, wherein is observed that it is possible to actualize wormholes that satisfy the Null Energy Condition (NEC) while operating within the framework of non-commutative geometry and modified gravity. A connection between the Casimir effect and noncommutative geometry is established in \cite{Kuhfittig:2023}, where it is found that the influence of noncommutative effects can be incorporated by modifying only the energy-momentum tensor in the Einstein field equations, while keeping the Einstein tensor unchanged. This approach enables the creation of macroscopic wormholes, despite the modest scale of the Casimir effect. Spherically symmetric wormhole solutions within the framework of modified gravity described by the function $f(R, T)$ is studied in \cite{Zubair:2017}. The authors introduced the widely recognized principles of non-commutative geometry, utilizing Gaussian and Lorentzian distributions of string theory. 

There is a recurring argument that modifications to General Relativity have the potential to elucidate the characteristics of the gravitational field at cosmic scales, eliminating the necessity to hypothesize the existence of an, as yet undetected, dark sector. Nevertheless, it's worth noting that the theoretical landscape in this regard appears to be considerably open-ended and unconstrained. The exploration of wormhole geometry, both within the framework of General Relativity and in modified theories, has always been an intriguing subject for researchers. Consequently, various methodologies and approaches have been employed to delve into the intricacies of wormhole geometry. Drawing inspiration from these diverse techniques utilized in modified theories, the authors in \cite{Shamir:2021} examined the wormhole solutions within the context of modified gravity theories. Traversable wormhole models within the framework of gravity theories by employing the Karmarkar condition are further explored in \cite{Malik:2022}. Their observation aligns with Einstein's field theory, which posits that wormholes require the existence of rare and unconventional material.

Another pivotal aspect of wormholes pertains to the violation of energy conditions within the framework of General Relativity (GR), particularly in the vicinity of the wormhole throat \cite{Morris:1988cz, Lobo:2007zb}. This necessitates the presence of a certain quantity of exotic matter, as the stress-energy tensor (SET) of matter violates the null energy condition (NEC), in order to keep the wormhole throat open. Consequently, the energy density of matter may be interpreted as negative, at least from certain reference frames. The pursuit of constructing wormholes with minimal reliance on exotic matter has garnered significant attention \cite{Visser:2003yf}. In their work \cite{Visser:2003yf}, the authors theoretically demonstrated that it is possible to minimize the amount of exotic matter and confine it to infinitesimal proportions at the wormhole throat through judicious selection of the wormhole's geometry. This mathematical approach is known as the "cut and paste" technique, and the resulting wormhole is referred to as a "thin-shell wormhole." The exploration of thin-shell wormholes has been documented in references \cite{Visser:1989kh,Visser:1989kg,Lobo:2003xd,Dias:2010uh}. Additionally, Nandi et al. \cite{Nandi:2004ku} have proposed an enhanced quantification method to precisely determine the quantity of exotic matter required in a given spacetime. It is demonstrated in \cite{Kuhfittig:2020} that non-commutative geometry, which is derived from string theory, can be considered a specific instance of $f(R)$ gravity. Consequently, it has the capacity to support a wormhole without necessitating the presence of exotic matter. Furthermore, it offers a rationale for selecting the $f(R)$ function within the modified gravitational theory. The authors in \cite{Baruah:2023} utilized non-commutative geometry as a gravitational source which is a viable approach in both GR and modified gravity for achieving complex space-time structures. Their examinations, conducted with rigorously constrained model parameters, reveal that it is feasible to construct wormholes conforming to the NEC within the context of non-commutative geometry when coupled with modified gravity.

Nevertheless, physicists continuously strive to either avoid energy condition violations or provide justifiable explanations for them. Unfortunately, up to this point, creating a static wormhole geometry that complies with energy conditions remains an elusive task. Consequently, researchers are exploring alternative strategies to address this challenge. This realization has sparked investigations into the potential existence of wormhole solutions within alternative theories of gravity, such as higher-order gravity theories \cite{Hochberg:1990is, Ghoroku:1992tz}, higher-dimensional cosmological wormholes \cite{Zangeneh:2014noa}, and the Einstein-Gauss-Bonnet theory \cite{Bhawal:1992sz, Maeda:2008nz, Mehdizadeh:2015jra}. In the realm of $f(R)$ gravity, it is theoretically feasible to construct traversable wormholes without relying on exotic matter \cite{Pavlovic:2014gba, Lobo:2009ip}, or by sourcing them with dark matter \cite{Muniz:2022eex}. Alternatively, researchers have explored the potential for wormholes in third-order Lovelock gravity \cite{KordZangeneh:2015dks, Mehdizadeh:2016nna}, hybrid metric-Palatini gravity \cite{Rosa:2021yym, KordZangeneh:2020ixt}, $f(Q)$ gravity \cite{Banerjee:2021mqk, Parsaei:2022wnu, Hassan:2022hcb}, and extended theories of gravity \cite{DeFalco:2021klh, DeFalco:2021ksd}. Additionally, researchers have examined traversable wormholes within the context of $f(R,T)$ gravity in previous studies \cite{Moraes:2017mir, Elizalde:2018frj, Moraes:2019pao}. Concurrently, authors in \cite{Zubair:2019uul, Rosa:2022osy, Banerjee:2020uyi} have uncovered exact wormhole solutions in $f(R,T)$ gravity that do not necessitate the presence of exotic matter. 

The potential existence of asymptotically flat wormhole configurations within the framework of Rastall-Rainbow modified gravity, which combines two distinct theoretical models: Rastall theory and the Rainbow description is observed in \cite{Tangphati:2023nwz}. Their investigation sheds light on how the interplay between Rastall parameters and Rainbow functions may mitigate violations of energy conditions within these modified gravity scenarios. In the subsequent sections, our focus will be on the examination of non-commutative effects on wormhole solutions within the recently proposed Rastall-Rainbow gravity theory \cite{Mota:2019zln}. The Rastall-Rainbow gravity theory, a synthesis of Rastall \cite{Rastall:1972swe} and Rainbow theories \cite{Magueijo:2002xx}, stands as an alternative approach to gravity.

Our plan for the present article is the following: In Section \ref{sec2}, we briefly review the newly proposed modified gravity theory, namely, the Rastall-Rainbow gravity. In the same Section, we describe the wormhole geometry and derive the field equations for traversable wormholes using static and spherically symmetric time-independent metric. Next, we  study the gravitational system in connection with noncommutative formulation of Rastall-Rainbow gravity, by adopting two different distributions of energy densities of a point-like gravitational sources in Section \ref{sec3} and Section \ref{sec4}, respectively. We assume  Gaussian and Lorentzian distribution separately and explore the properties of wormholes in those Sections. Finally, in Section \ref{sec5}, we discuss our results and findings.

%-----------------------------------------------------

\section{Rastall-Rainbow gravity and stellar structure equations}\label{sec2}

\subsection{Rastall-Rainbow theory}
\label{TOV_RR}

The Rastall-Rainbow gravity model, as described in \cite{Mota:2019zln}, offers a coherent fusion of two modified gravity theories that expand upon General Relativity (GR): the Rastall theory \cite{Rastall:1972swe} and Rainbow gravity \cite{Magueijo:2002xx}. The genesis of the latter theory traces back to 2004, when Jo\~ao Magueijo and Lee Smolin introduced a remarkable extension of nonlinear special relativity into the realm of curved spacetime. Through adjustments to the formalism governing the tenets of this relativity, they successfully accommodated curvature, leading to the emergence of the concept referred to as 'double general relativity.' This innovative approach brought forth a fresh perspective on the nature of spacetime and its dynamic properties.

A significant outcome of this theoretical investigation was the unveiling of 'Rainbow gravity,' which alters the conventional relativistic dispersion relation, $E^{2} - p^{2} = m^{2}$, particularly in the high-energy regime. This modification introduces two flexible functions, known as 'Rainbow functions,' denoted as $\Xi(x)$ and $\Sigma(x)$, as expressed by the following equation:
\begin{equation}
E^{2} \Xi(x)^{2} - p^{2}\Sigma(x)^{2} = m^{2}.
\label{eq1}
\end{equation}

In this expression, the symbol $x = E/E_{p}$ signifies the dimensionless ratio between the energy of the test particle, denoted as $E$, and a critical energy often regarded as the Planck energy, represented as $E_{p} = \sqrt{\frac{\hslash c^{5}}{G}}$. This energy quantity holds fundamental significance in the realm of physics, serving as a defining scale where quantum gravitational effects become notably significant. Consequently, the functions $\Xi(x)$ and $\Sigma(x)$, with specific functional forms inspired by high-energy phenomena, assume a central role within the framework of Rainbow gravity. They introduce a profound dependence on the energy of the test particle into the geometry of spacetime. As a result, when particles approach extreme energies near the Planck scale, their motion is substantially influenced by these energy-dependent metrics, giving rise to significant effects in the behavior of spacetime that resemble backreaction phenomena. Conversely, in the low energy approximation where $x = E/E_{p} \rightarrow 0$, the functions $\Xi(x)$ and $\Sigma(x)$ are tailored in such a way that the usual dispersion relation is reinstated, satisfying the following relationships:
\begin{equation}
    \lim_{x\rightarrow 0} \Xi(x)=1, \quad \lim_{x \rightarrow 0} \Sigma(x)=1.
    \label{eq2}
\end{equation}

In this context, the description of spacetime employs a metric that depends on energy, as outlined in \cite{Magueijo:2002xx}, and is expressed as:
\begin{equation}
g^{\mu\nu}(x)=\eta^{ab} e_{a}^{\mu}(x)\otimes e_{b}^{\nu}(x),
\label{eq3}
\end{equation}
where the energy-dependent vierbein fields represented as $e_{a}^{\mu}(x)$, are connected to the independent vierbein fields denoted by $\widetilde{e}_{a}^{\mu}$ through the following relationships:
\begin{equation}
e_{0}^{\mu}(x)=\frac{1}{\Xi(x)} \widetilde{e}_{0}^{\mu}, \quad e_{k}^{\mu}(x)=\frac{1}{\Sigma(x)} \widetilde{e}_{k}^{\mu}.
\label{eq4}
\end{equation}
In this context, the index $k$ is used to denote the spatial coordinates, typically taking values from the set $(1, 2, 3)$. This modification in the description of spacetime geometry is a fundamental feature of Rainbow gravity, wherein the core premise is that the spacetime's geometry is contingent upon the energy of the test particle (EPT). Consequently, all the quantities that play a role in the field equations within this gravitational theory acquire energy-dependent characteristics. In contrast to the conventional Einstein's field equations, Rainbow gravity introduces a collection of alternative field equations. One illustrative example of such an equation is:
\begin{equation}
G_{\mu\nu}(x) \equiv R_{\mu\nu}(x) - \frac{1}{2}g_{\mu\nu}(x)R(x) = k(x)T_{\mu\nu}(x),
\label{eq10}
\end{equation}
where $k(x) = 8 \pi G(x)$.
These alterations hold considerable significance and provide fresh perspectives on the intricate connections between gravity, high-energy physics, and the fundamental fabric of spacetime.

Peter Rastall made a notable contribution to gravitational theory in the early 1970s when he proposed a modification to the conservation principles governing the energy-momentum tensor in curved spacetime, as detailed in \cite{Rastall:1972swe}. His pioneering work aimed to preserve the integrity of the gravitational Bianchi identity of the Einstein geometric tensor while introducing necessary adjustments. Rastall's concept revolved around the idea that the divergence of the energy-momentum tensor ($T_{\mu\nu}$) was directly proportional to the variation of the Ricci scalar ($R$), implying that the geometry itself must contribute to the total energy of the system. While certain arguments have been put forth suggesting that Rastall gravity is entirely equivalent to the standard Einstein gravity \cite{Visser:2017gpz}, the validity of this assertion has been subjected to scrutiny by various studies \cite{Darabi:2017coc}.

The modified conservation law introduced by Rastall is formulated in \cite{Mota:2019zln}: 
\begin{equation}
\nabla^{\mu}T_{\mu\nu} = \bar{\lambda}\nabla_{\nu}R,
\label{eq6}
\end{equation}
where $\bar{\lambda}$ is defined as $\frac{1 - \lambda}{16 \pi G}$, with $\lambda$ denoting the Rastall parameter that characterizes the coupling between geometry and matter fields, as outlined in \cite{Das:2018dzp}. When $\bar{\lambda} = 0$ (or $\lambda = 1$), General Relativity (GR) is retrieved. In flat spacetime, where the Ricci scalar $R$ becomes zero, the conventional conservation law for $T_{\mu\nu}$ is reinstated. Consequently, the modified Einstein field equations consistent with equation (\ref{eq6}) can be expressed as provided in \cite{Mota:2019zln}: 
\begin{equation}
R_{\mu}{}^{\nu} - \frac{\lambda}{2}\delta_{\mu}{}^{\nu}R = 8\pi  T_{\mu}{}^{\nu},
\label{eq8}
\end{equation}

The Rastall-Rainbow model combines the influences of both Rainbow gravity and Rastall gravity within a unified framework. The field equations within this consolidated formalism are expressed as follows:
\begin{equation}
R_{\mu}{}^{\nu}(x) - \frac{\lambda}{2}\delta_{\mu}{}^{\nu}(x)R(x) = k(x) T_{\mu}{}^{\nu}(x)  
 \label{eq7},
\end{equation}
In this context, the Rastall parameter $\lambda$ remains constant and is not influenced by the energy of the test particle. In the ensuing discussion, our objective is to describe the most basic form of wormhole geometry and examine its geometric characteristics.

\subsection{The wormhole geometry and the field equations}

The present study aims to explore asymptotically flat wormhole geometries in Rastall-Rainbow gravity. For this purpose,
we consider a static and spherically symmetric line element and investigate the effects of model parameters. Thus, we 
usurp the usual GR quantities $\widetilde{e}_{i}$ for spherical symmetry into Eq. (\ref{eq3}), which gives
\begin{equation} 
    ds^{2}=-\frac{e^{2\Phi(r)}}{\Xi^{2}(x)} dt^{2}+ \frac{dr^2}{\Sigma^{2}(x) \left(1-\frac{b(r)}{r}\right)}+\frac{r^{2}}{\Sigma^{2}(x)}(d\theta^{2}+\sin{\theta}^{2}d\phi^{2}),
    \label{eq5}
\end{equation}
where $\Phi(r)$ is denoted the redshift function and  $b(r)$ is the shape function that determines the shape of the wormhole.
Interestingly, the rainbow functions $\Xi(x)$ and $\Sigma(x)$ are attached with metric potentials, but the spherical coordinate
($r$, $t$, $\theta$,  $\phi$) are independent of the energy probe particles in gravity's rainbow.  For the wormhole to be traversable,
the redshift function $\Phi(r)$ must be finite everywhere which ensure the  horizonless spacetime. On the other hand, the shape function $b(r)$ should satisfy the flaring-out condition,  i.e., $\frac{b(r)-rb^{\prime}(r)}{b^2(r)}>0$ \cite{Morris:1988cz}.  More precisely, it should  satisfy the condition  $b^{\prime}(r_0) < 1$. More precisely, it should  satisfy the condition  $b^{\prime}(r_0) < 1$, where $b(r_0)= r_0$ is the  minimum value of surface radius called the wormhole throat with proper circumference $2\pi r$. In addition to this $b(r)$ should be defined as the positive function and $1-b(r)/r > 0$ for the region out of the throat.

For a complete description of the configuration,
we assume the energy-momentum tensor of the anisotropic fluid distribution, which can write in the following form
\begin{equation}\label{eq12}
T_{\mu\nu}=(\rho+p_t)u_\mu u_\nu+ p_t g_{\mu\nu}-(p_{t}-p_{r}) \chi_{\mu}\chi_{\nu},
\end{equation}
where $u^\mu$ is the fluid 4-velocity ($u_\mu u^\mu = -1$), $\chi_{\mu}$ is the unit radial vector so that $\chi_{\mu} \chi^{\mu} = 1$. Whereas the quantities, $\rho = \rho(r)$ represents the energy density, $p_r = p_r(r)$ is the radial pressure and $p_t = p_t (r)$ is the transverse pressure, respectively.

Following the paper \cite{Mota:2019zln}, one can rewriting the Eq. (\ref{eq8}) in its covariant form, and obtain the modified Einstein's field equations:
\begin{equation}
     R_{\mu\nu}-\frac{1}{2}g_{\mu\nu}R=8\pi \left[T_{\mu\nu}-\frac{(1-\lambda)}{2(1-\lambda)}g_{\mu\nu}T\right] .
     \label{eq13b}
 \end{equation}
 As one can see from the above equation that the Einstein tensor on the left-hand side and effective energy-momentum tensor on the right-hand side. Finally, the nonzero components of equation of motion (\ref{eq13b}) are given by \cite{Tangphati:2023nwz}
  \begin{eqnarray}
&& \frac{b^{\prime}}{r^{2}} = 8\pi  \Bar{\rho},  \label{eq14} \\
&& 2\left(1-\frac{b}{r}\right)\frac{\Phi^{\prime}}{r}-\frac{b}{r^{3}}  = 8\pi  \Bar{p}_{r},  \label{eq15} \\
&& \left(1-\frac{b}{r}\right)\left[\Phi^{\prime\prime}+\Phi^{\prime 2}-\frac{b^{\prime}r-b}{2r(r-b)}\Phi^{\prime}-\frac{b^{\prime}r-b}{2r^2(r-b)} +\frac{\Phi^{\prime}}{r}\right] 
 = 8\pi  \Bar{p}_{t},  \label{eq16}
\end{eqnarray}
where $\Bar{\rho}$ represents the effective energy density, $\Bar{p}_{r}$ is the effective radial pressure and $\Bar{p}_{t}$ is the effective tangential pressure respectively, we define
\begin{align}
    \Bar{\rho} & = \frac{1}{\Sigma(x)^{2}}\left[\alpha_{1}\rho+\alpha_{2}p_{r}+2\alpha_{2}p_{t}\right],\label{eq17}\\
    \Bar{p}_{r} & = \frac{1}{\Sigma(x)^{2}}\left[\alpha_{2}\rho+\alpha_{1}p_{r}-2\alpha_{2}p_{t}\right], \label{eq18}\\
    \Bar{p}_{t} & = \frac{1}{\Sigma(x)^{2}}\left[\alpha_{2}\rho-\alpha_{2}p_{r}+\alpha_{3}p_{t}\right], \label{eq18b}
\end{align}
with
\begin{equation*}
    \alpha_{1}=\frac{1-3\lambda}{2(1-2\lambda)}, \qquad \alpha_{2}=\frac{1-\lambda}{2(1-2\lambda)}, \qquad
    \alpha_{3}=-\frac{\lambda}{1-2\lambda}.
\end{equation*}
In addition, we see that field equations (\ref{eq14})$-$(\ref{eq16}) is different from Einstein field equation (EFE), but one can recover EFE when $\lambda=1$ and $\Sigma=1$. Since, the Eqs. (\ref{eq14})$-$(\ref{eq16})  provide three independent equations, with five unknown quantities, i.e., $\Phi(r)$, $b(r)$, $\rho(r)$, $p_r(r)$ and $p_t(r)$, respectively. Considering different strategies
one can reduce the number of unknown functions for complete description of the wormhole geometry. Due to the complexity of the above equations, we focus on the simplest assumption 
with a constant redshift function i.e., $\Phi'=0$, and the effect of noncommutativity via  two different distributions of energy density (Gaussian and Lorentzian). Finally, we would like to recall the conservation equation of the energy-momentum tensor $T^{\nu}_{\:\:\:\mu;\nu}=\Bar{\lambda}R_{\mu}$  leads to
the following relation,
\begin{equation}
 \Bar{p}_{r}' = -(\Bar{p}_{r}+\Bar{\rho})\Phi'+ \frac{2}{r}\left(\Bar{p}_{t}-\Bar{p}_{r}\right).
 \label{eq21}
\end{equation}

%%%%%%%%%%%%%%%%%%%%%%%%%%%%%%%%%%%%%%%%%%%%%%%%%%%%%%%%%%%%%%%%%%%%%%%%%%%%%%%%%%%%%%%%%%%%

\section{ Gaussian distribution} \label{sec3}

\begin{figure}[h]
    \centering
    \includegraphics[width = 7 cm, height=7cm]{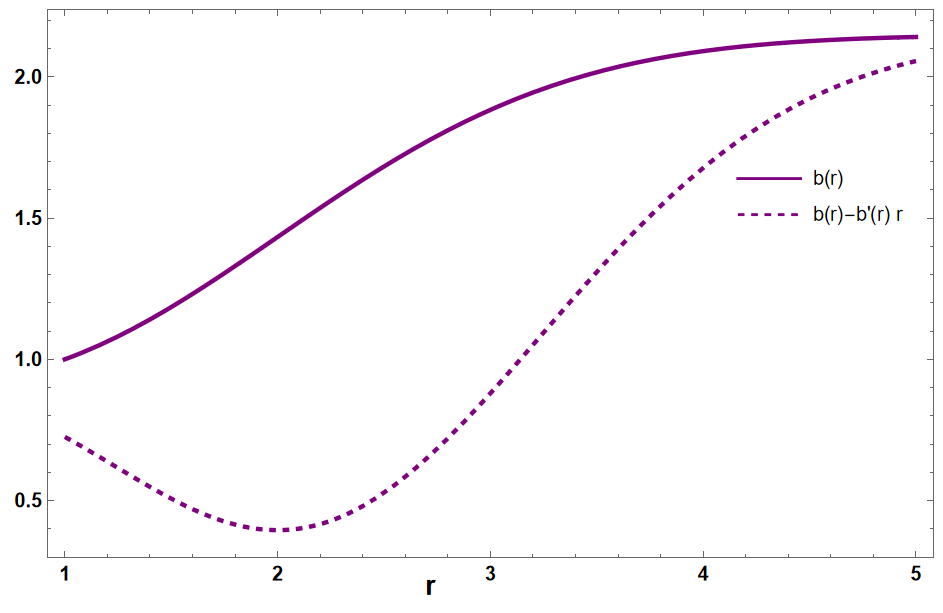}
    \includegraphics[width = 7.2 cm, height=7.2cm]{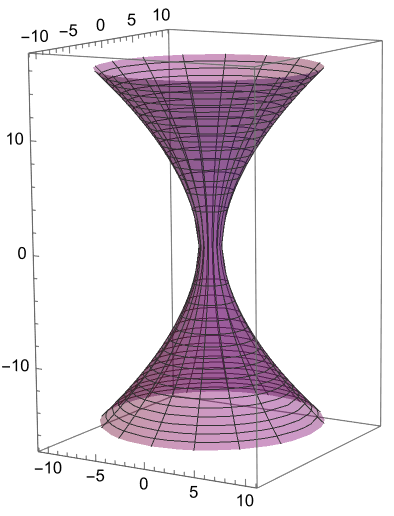}
    \caption{ Left figure: Plot for $b(r)$ and $b(r)-b'(r) r$ using the Eq. (\ref{sh1}). Right figure: The embedding diagram for a wormhole. In both plots we have taken $r_0= 1$, $M= 1$, $\alpha = 1$, $\Sigma = 2$ and $\lambda =0.4$, respectively. }
    \label{fig:1}
\end{figure}

Here we study the above gravitational system in connection with noncommutative formulation of Rastall-Rainbow gravity. Our starting point is to consider the energy density of a point-like gravitational source with a Gaussian distribution,
\begin{eqnarray}\label{GD}
 \rho (r)=\frac{M }{(4 \pi  \alpha )^{3/2}}   \exp \left(-\frac{r^2}{4 \alpha }\right),
\end{eqnarray}
where $\alpha$ represents the noncommutative parameter and $M$ is the  mass diffused throughout a region with linear size $\sqrt{\alpha}$. A proper justification has  been given in Ref. \cite{Nicolini:2005vd} for choosing the above form for the energy density. Now, using equations (\ref{eq14}-\ref{eq16}) and Eq. (\ref{GD}) for zero-tidal-force wormholes, we obtain the shape function, 
\begin{eqnarray}\label{sh1}
 b(r) =\frac{2 M} {\lambda  \Sigma^2}\left(\text{erf}\left(\frac{r}{2 \sqrt{\alpha }}\right)-\frac{r e^{-\frac{r^2}{4 \alpha }}}{\sqrt{\pi \alpha}}\right)+c_1,
\end{eqnarray}
where $c_1$ is the integrating constant. This value can be determined using the condition $b(r_0)=r_0$, which yield
\begin{eqnarray}\label{sh1}
 b(r) = r_0+ \frac{2 M} {\lambda  \Sigma^2}\left(\text{erf}\left(\frac{r}{2 \sqrt{\alpha }}\right)-\frac{r e^{-\frac{r^2}{4 \alpha }}}{\sqrt{\pi \alpha}}\right)-\frac{2 M} {\lambda  \Sigma^2}\left(\text{erf}\left(\frac{r_0}{2 \sqrt{\alpha }}\right)-\frac{r_0 e^{-\frac{r^2_0}{4 \alpha }}}{\sqrt{\pi \alpha}}\right) . 
\end{eqnarray}
The shape function depends on the both parameters $\lambda $ and $\Sigma$, respectively. Thus, in order to satisfy the necessary conditions for a static wormhole solutions, we consider a particular wormhole model with the throat at $r_0=1$ and $b'(r_0=1)= \frac{M r_0^2 }{\sqrt{\pi } \alpha ^{3/2} \lambda  \Sigma ^2} e^{-\frac{r_0^2}{4 \alpha }} \thickapprox 0.49 <1$. This can be achieved by fixing the other parameters $M= 1$, $\alpha = 1$, $\Sigma = 1.5$ and $\lambda = 0.4$, respectively. With this assumption, we plot the shape function and the embedding diagram through a $2\pi$ rotation around the $z$-axis that provides valuable insights into the wormhole's existence, see Fig. \ref{fig:1}.

\begin{figure}[h]
    \centering
    \includegraphics[width = 8 cm, height=7cm]{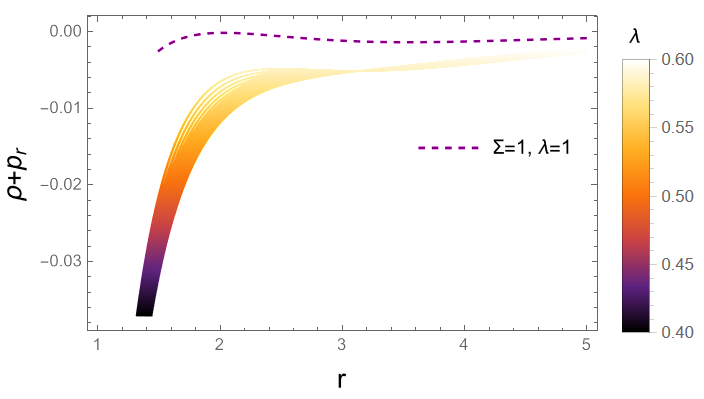}
    \includegraphics[width = 8 cm, height=7cm]{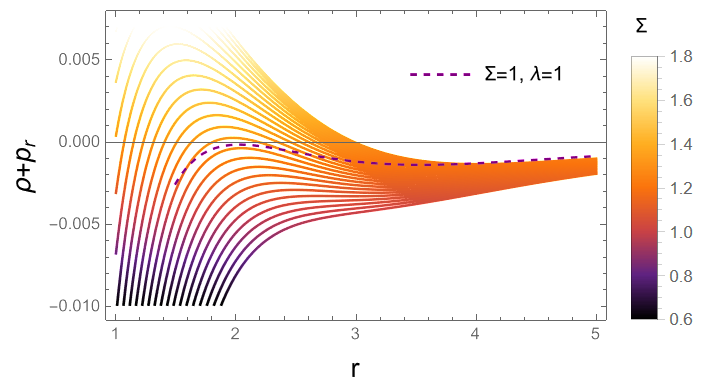}
    \caption{Plots for $\rho+p_r$ with varying the parameter $\lambda \in [0.36, 0.6]$ while keeping $\Sigma=2$ fixed, and then we vary $\Sigma \in [0.6, 1.8]$ while keeping $\lambda=0.4$ fixed. Other parameters are same as of Fig. \ref{fig:1}. The dashed line represents 
    the GR solution for noncommutative geometry for $\lambda =1$ and  $\Sigma =1$.}
    \label{fig:2}
\end{figure}

\begin{figure}[h]
    \centering
    \includegraphics[width = 8 cm, height=7cm]{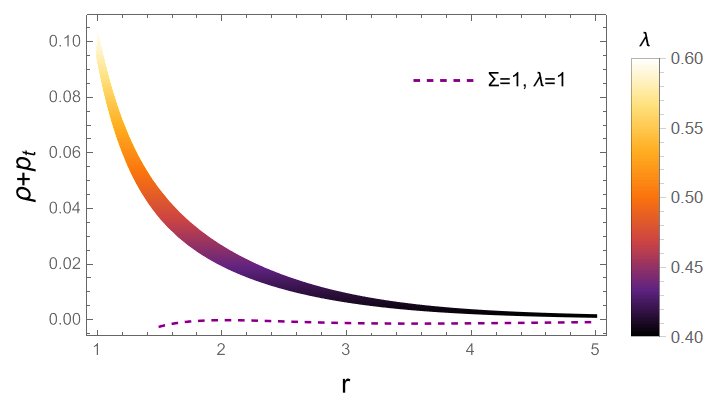}
    \includegraphics[width = 8 cm, height=7cm]{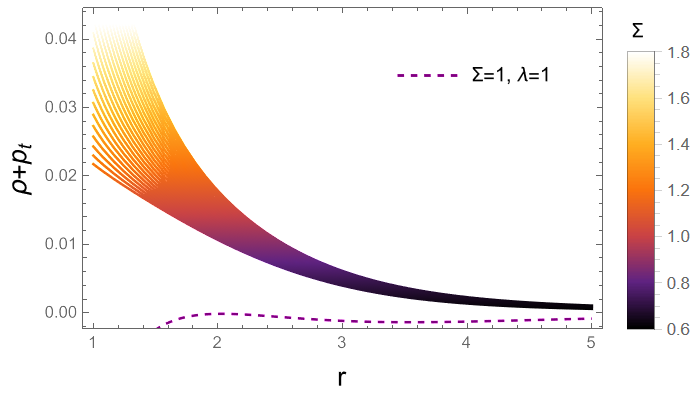}
    \caption{Plots for $\rho+p_t$ with varying the parameter $\lambda \in [0.36, 0.6]$  while keeping $\Sigma=2$ fixed, and then we vary $\Sigma \in [0.6, 1.8] $ while keeping $\lambda=0.4$ fixed. Other parameters are same as of Fig. \ref{fig:1}. The dashed line represents 
    the GR solution for noncommutative geometry for $\lambda =1$ and  $\Sigma =1$.}
    \label{fig:3}
\end{figure}

With the foregoing assumptions, the other two nonvanishing components of the gravity field equations are given by
 \begin{eqnarray}
    p_r(r)&=&-\frac{\Sigma^2 \left((\lambda -1) r b'(r)+b(r)\right)}{8 \pi  r^3},\label{eq25}\\
    p_t(r)&=&\frac{\Sigma^2 \left((1-2 \lambda ) r b'(r)+b(r)\right)}{16 \pi  r^3}\label{eq26},
\end{eqnarray}
where dashes denote derivatives with respect to $r$, and one can easily get the full expression using the Eq. (\ref{sh1}). Let us begin by examining the energy conditions associated with the wormhole matter by using the field equations (\ref{eq25})-(\ref{eq26}) and (\ref{GD}). Here, we restrict our analysis to the null energy condition (NEC), which is define by $\rho+p_{r} \geq 0$ and $\rho+p_{t} \geq 0$. We start our construction by analyzing separately the parameters  based on values used in Rastall-Rainbow gravity. Here, we vary just the $\lambda$ parameter while keeping $\Sigma$ fixed, and then we vary $\Sigma$ while keeping $\lambda$ fixed. In Fig. \ref{fig:2}, we plot for $\rho+p_r$ by setting the parameters: $r_0= 1$, $M= 1$, $\alpha = 1$, $\Sigma = 2$ and varying $\lambda \in [0.36, 0.6]$ and  $\Sigma \in [0.6, 1.8]$, selectively. We can observe that $\rho+p_r<0$ in both cases. Note that $\lambda =1$ and  $\Sigma =1$ represents the GR solution for noncommutative geometry. Next, we plot for $\rho+p_t$ considering the same sets of parameters in Fig.  \ref{fig:3}. It turns out that $\rho+p_t>0$ throughout the spacetime. We thus find that the NEC is violated, due to the flaring-out condition. Using the same expansion, we finally obtain NEC of the wormhole solution at the throat, 
 \begin{eqnarray}
    (\rho+p_r)|_{r_0}&=&\frac{M e^{-\frac{r_0^2}{4 \alpha }}}{8 \pi ^{3/2} \alpha ^{3/2} \lambda }-\frac{\Sigma ^2}{8 \pi  r_0^2},\label{pro}\\
   (\rho+ p_t)|_{r_0} &=&\frac{\frac{M e^{-\frac{r_0^2}{4 \alpha }}}{\alpha ^{3/2} \lambda }+\frac{\sqrt{\pi } \Sigma ^2}{r_0^2}}{16 \pi ^{3/2}}\label{pto}.
\end{eqnarray}
Also in this case, we found that $(\rho+p_r)|_{r_0}<0$ and $(\rho+p_t)|_{r_0}>0$. Thus, in the vicinity of the wormhole throat the NEC is violated within the specified range. 

Let us now focus our attention on the stability analysis of the above static solutions based on the adiabatic sound velocity, denoted as $v_s^2= \frac{\partial{<p>}}{\partial{\rho}}$, where $<p>$ represents the average pressure across the three spatial dimensions, namely $<p> = \frac{1}{3}(p_r + 2p_t)$. This property is valid under the constraint $0 \leq v_s^2 < 1$. Now by substituting Eq. (\ref{GD}) and Eqs. (\ref{sh1})-(\ref{eq25}), we find
\begin{equation}\label{Sound}
v_s^2 = \frac{d<p>}{dr} \left( \frac{d\rho}{dr}\right)^{-1} = \frac{2}{3 \lambda} - 1.
\end{equation}
Interestingly, the sound velocity depends only on the Rastall parameter $\lambda$, and to satisfy the compactified space coordinate within the interval $\left(\frac{1}{3}, \frac{2}{3}\right)$. We perform all the above calculation keeping in mind the range of $\lambda$, and see that the flaring-out and stability conditions are simultaneously obeyed.

\begin{figure}[h]
    \centering
    \includegraphics[width = 8 cm, height=6cm]{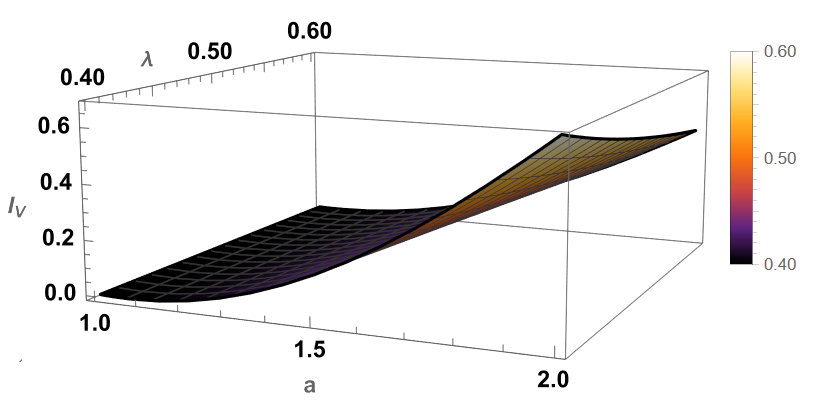}
    \includegraphics[width = 8 cm, height=6cm]{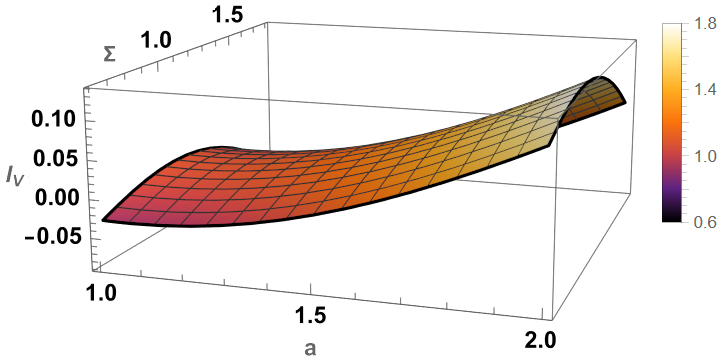}
    \caption{We plot the total amount of exotic matter specifying a certain volume that needed to construct for wormhole solution. See the text for details.}
    \label{fig:7}
\end{figure}

To maintain the stability and flare-out condition, we need an amount of exotic matter, and quantifying the amount of exotic matter (NEC violation) we consider the \enquote{volume integral quantifier} \cite{Kar:2004hc}. Here, we shall evaluate the
total amount of exotic matter for the constructed wormhole by the definite integral (with a cut-off of the stress-energy at $a>r_0$): 
\begin{equation}
I_V = \int\left(\rho+p_r\right)dV= 2\int^{a}_{r_0}\left(\rho+p_r\right)4\pi r^2 dr.
\end{equation}
By taking into account (\ref{GD}) and (\ref{eq25}), and using (\ref{sh1}), we have (for simplicity, we consider $M=1$ and $\alpha=1$ as of Fig. \ref{fig:1})
\begin{equation}\label{VIQ}
I_V = \frac{1}{2 \sqrt{\pi} \lambda  } \left[2 \sqrt{\pi } \left(\text{erf}\left(\frac{r}{2}\right)-\text{erf}\left(\frac{r_0}{2}\right)\right)+r E_{\frac{3}{2}}\left(\frac{r^2}{4}\right)-4 e^{-\frac{r^2}{4}} r+2 e^{-\frac{r_0^2}{4}} r_0+\sqrt{\pi } \lambda  r_0 \Sigma ^2\right]_{r_0}^{a}.
\end{equation}
The essence of this discussion is when the limit $a\to r_0$, one verifies that $I_V = \int\left(\rho+p_r\right)dV \to 0$. 
To describe the difficult task, we plot Fig. \ref{fig:7}. Now, it is clear from Fig. \ref{fig:7} that  one may construct wormholes in Rastall-Rainbow gravity with arbitrarily small amounts of NEC violating matter. We have considered the same sets of parameters as of Figs. \ref{fig:2}
and \ref{fig:3}.

%%%%%%%%%%%%%%%%%%%%%%%%%%%%%%%%%%%%%%%%%%%%%%%%%%%%%%%%
\section{Lorentzian distribution} \label{sec4}

\begin{figure}[h]
    \centering
    \includegraphics[width = 7 cm, height=7cm]{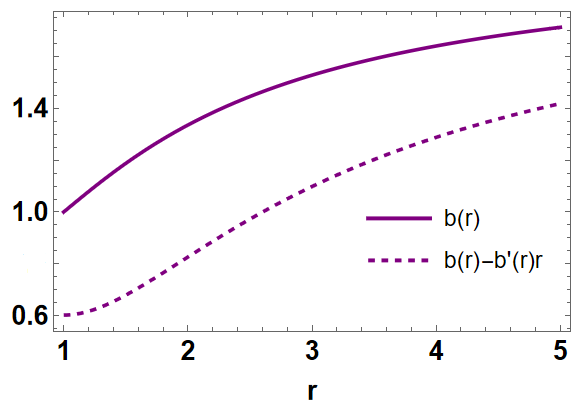}
    \includegraphics[width = 7.2 cm, height=7.5cm]{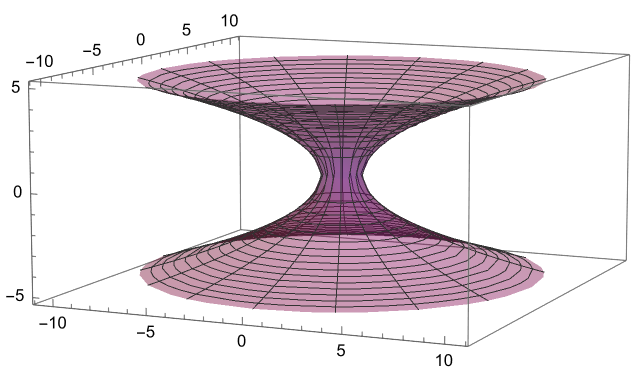}
    \caption{ Left figure: Plot for $b(r)$ and $b(r)-b'(r) r$ using the Eq. (\ref{sh2}). Right figure: The embedding diagram for a wormhole. In both plots we have taken $r_0= 1$, $M= 1$, $\Theta = 1$, $\Sigma = 2$ and $\lambda = 0.4$, respectively. }
    \label{fig:4}
\end{figure}

Here, we shall replace the Gaussian distribution functions with the following Lorentzians distribution.
Within noncommutative geometry and using the Lorentzians distribution, we explore the properties of wormholes.  We start with the following definition, 
\begin{eqnarray}\label{LD}
 \rho (r)=\frac{M }{ \pi^{2}} \left(\frac{\sqrt{ \Theta}}{( r^2+\Theta)^2 }\right),
\end{eqnarray}
where $\Theta$ is the strength of non-commutativity of spacetime and $M$ is the  mass diffused throughout a region with linear size $\sqrt{\Theta}$. Using  equations (\ref{eq14}-\ref{eq16}) and Eq. (\ref{GD}), we get the following shape function for zero-tidal-force wormholes,
\begin{eqnarray}\label{LD}
b(r)=  \frac{4 M \left(\left(\Theta +r^2\right) \tan ^{-1}\left(\frac{r}{\sqrt{\Theta }}\right)-\sqrt{\Theta } r\right)}{\pi  \lambda  \left(\Theta +r^2\right) \Sigma (x)^2}+c_2,
\end{eqnarray}
where the integrating constant $c_2$ can be determined from the condition $b(r_0)=r_0$, and one gets
\begin{eqnarray}\label{sh2}
b(r)= r_0+ \frac{4 M }{\pi  \lambda \Sigma^2} \left(\tan ^{-1}\left(\frac{r}{\sqrt{\Theta }}\right)-\frac{\sqrt{\Theta } r}{\left(\Theta +r^2\right)}\right)- \frac{4 M }{\pi  \lambda \Sigma^2} \left(\tan ^{-1}\left(\frac{r_0}{\sqrt{\Theta }}\right)-\frac{\sqrt{\Theta } r_0}{\left(\Theta +r_0^2\right)}\right).
\end{eqnarray}
As of the same in previous case the shape function depends on the both parameters $\lambda $ and $\Sigma$. The next step is to investigate the protecting property of wormholes, where we consider a particular model with the throat radius at $r_0=1$ and $b'(r_0=1)= \frac{8 \sqrt{\Theta } M r_0^2}{\pi  \lambda  \Sigma ^2 \left(\Theta +r_0^2\right)^2} \thickapprox 0.16 <1$.  We use the following set of values: $M= 1$, $\Theta = 1$, $\Sigma = 2$ and $\lambda = 1$, respectively. In Fig. \ref{fig:4}, we show the shape function and the embedding diagram using the same sets of parameters. 

\begin{figure}[h]
    \centering
    \includegraphics[width = 8.3 cm, height=7cm]{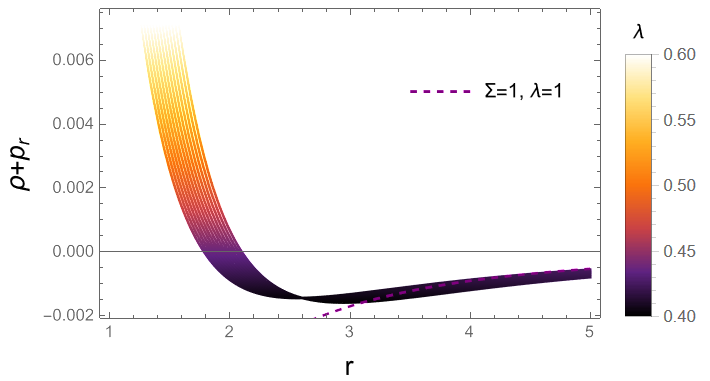}
    \includegraphics[width = 8.3 cm, height=7cm]{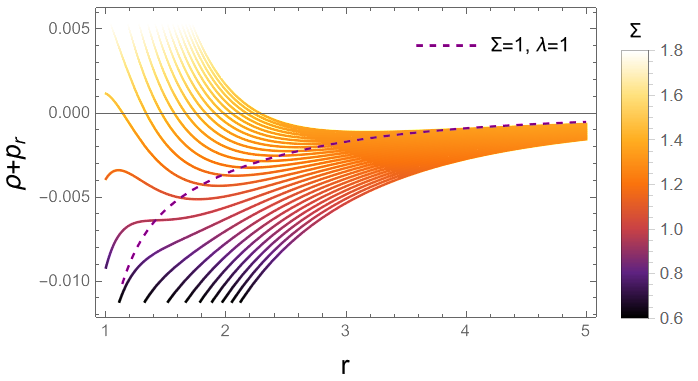}
    \caption{Plots for $\rho+p_r$ with varying the $\lambda \in [0.36, 0.6]$ parameter while keeping $\Sigma=0.8$ fixed, and then we vary $\Sigma \in [0.6, 1.8]$ while keeping $\lambda=0.4$ fixed. Other parameters are same as of Fig. \ref{fig:1}. The dashed line represents 
    the GR solution for noncommutative geometry with $\lambda =1$ and  $\Sigma =1$.}
    \label{fig:5}
\end{figure}

\begin{figure}[h]
    \centering
    \includegraphics[width = 8.3 cm, height=7cm]{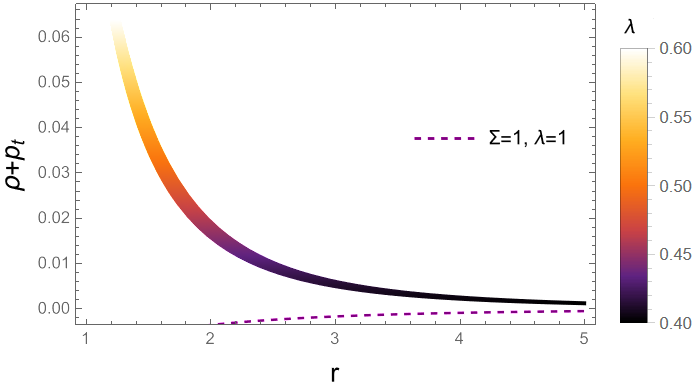}
    \includegraphics[width = 8.3 cm, height=7cm]{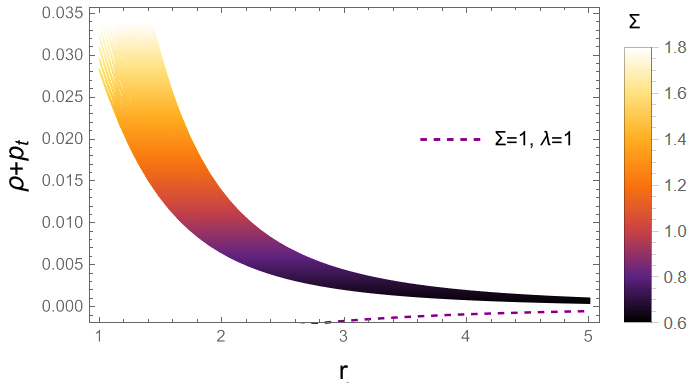}
    \caption{Plots for $\rho+p_t$ with varying the $\lambda \in [0.36, 0.6]$ parameter while keeping $\Sigma=0.8$ fixed, and then we vary $\Sigma \in [0.6, 1.8]$ while keeping $\lambda=0.4$ fixed. Other parameters are same as of Fig. \ref{fig:1}. The dashed line represents 
    the GR solution for noncommutative geometry with $\lambda =1$ and  $\Sigma =1$.}
    \label{fig:6}
\end{figure}

Next, we  explore the NEC along radial and transverse pressures respectively.  Now, substituting Eq. (\ref{sh2}) into  the Eqs. (\ref{LD}) and (\ref{eq25})-(\ref{eq26}), we get the full expressions for stress-energy components.  The qualitative behaviour of $\rho+p_r$ and $\rho+p_t$ are depicted in Figs. \ref{fig:5} and \ref{fig:6}, respectively. Imposing different values on $\lambda \in [0.36, 0.6]$ and $\Sigma \in [0.6, 1.8]$,  we plot the quantities for $\rho+p_r$ in Fig. \ref{fig:5}.  We see form figures that the NEC is satisfying in the vicinity of the wormhole throat for higher values of $\Sigma$, but for lower values of $\Sigma$,  NEC is violated. In Fig. \ref{fig:6}, we depict $\rho+p_t$ for varying $\lambda$ and $\Sigma $, respectively. It is clear from the figures that $\rho+p_t>0$  throughout the spacetime for any values of Rastall-Rainbow parameters. At the throat, this reduces to 
 \begin{eqnarray}
    (\rho+p_r)|_{r_0}&=& \frac{\sqrt{\Theta } M}{\pi ^2 \lambda  \left(\Theta +r_0^2\right)^2}-\frac{\Sigma ^2}{8 \pi  r_0^2},\label{pro1}\\
   (\rho+ p_t)|_{r_0} &=& \frac{1}{16 \pi ^2}\left(\frac{8 \sqrt{\Theta } M}{\lambda  \left(\Theta +r_0^2\right)^2}+\frac{\pi  \Sigma ^2}{r_0^2}\right), \label{pto2}
\end{eqnarray}
Taking into account the condition $b'(r_0)<1 $ and $\frac{\sqrt{\Theta } M}{\pi ^2 \lambda  \left(\Theta +r_0^2\right)^2}>\frac{\Sigma ^2}{8 \pi  r_0^2}$, we have  always $(\rho+p_r)|_{r_0}>0$ and $\rho+p_t>0$. With this consideration we have that always $\rho+p_{r,t}>0$, which allows us to conclude that the NEC is satisfied at the wormhole throat. Moreover, we can see that energy density (\ref{LD}) under consideration is positive throughout spacetime. Thus, the assumption of Lorentzian distributed for a static spacetime entails that the WEC and NEC are both
satisfied at the throat and its neighbourhood, contrary to
their GR counterparts.

Also we discuss the possible stable region for wormhole geometry through the adiabatic sound velocity, using the Eq. (\ref{sh2}) into  the Eqs. (\ref{LD}) and (\ref{eq25})-(\ref{eq26}), we have
\begin{equation}\label{Sound3}
v_s^2 = \frac{d<p>}{dr} \left( \frac{d\rho}{dr}\right)^{-1} = \frac{2}{3 \lambda }-1.
\end{equation} 
Interestingly, the adiabatic sound velocity is identical with the previously analyzed cases where the stable region is found 
within the interval $\left(\frac{1}{3}, \frac{2}{3}\right)$. As we can see from the above analysis that the WEC is satisfied at the throat and its neighbourhood depending on the suitable choice of Rastall-Rainbow parameters.

%%%%%%%%%%%%%%%%%%%%%%%%%%%%%%%%%%%%%%%%%%%%%%%%%%%%%%%%%%%%

\section{Concluding remarks }\label{sec5}

Recently, a new alternative gravity theory was proposed in \cite{Mota:2019zln} as a combination of two theories namely the Rastall theory and the Rainbow theory and dubbed as Rastall-Rainbow gravity, which relaxes the conservation of energy-momentum. It was pointed out that the spherically symmetric metric
is dependent on the energy of the probe particles through the rainbow functions. Inspired by this new proposal of Rastall-Rainbow
theory, in this work, we explore the existence of traversable wormhole geometries and studied their most important properties and features. 

For this purpose, we start with the assumption of static and spherically symmetric metric and deduced the modified field equations. With these field equations, we presented the complete analytical solution for a restricted class of wormhole geometries by adopting  two different distributions of energy density (Gaussian and Lorentzian).  But our primary interest is to analyze the effect of the Rastall and the Rainbow parameters on the  wormhole geometries. To put constraints on those parameters, we perform the adiabatic  sound velocity and determine the region of stable wormhole solution. Interestingly, the results show that the sound speed  depends only on the Rastall parameter.  

Furthermore, we have studied the energy conditions especially focus on NEC along radial and transverse directions depending on both parameters. Because the violation of NEC is a fundamental property of wormhole physics in classical GR. We have presented the behavior of quantities by plotting graphs for different values of suitable parameters. We found out that the NEC is violated for Gaussian distribution regardless of the values of $\lambda$ and $\Sigma$, which characterized the solutions. However, in principle it is possible to find solutions satisfying NEC for Lorentzian distribution at the throat and its neighbourhood, contrary to their GR counterparts. Finally, we conclude that obtained solution is interesting which can be useful in exploring a suitable model for wormhole.
 
\section*{date availability}

There are no new data associated with this article.

\begin{acknowledgments}
In accordance with the visiting associateship scheme, A. Pradhan is grateful to IUCAA, Pune, India, for providing support and facilities.
\end{acknowledgments}\

\end{document}